\documentclass{PoSa}
\usepackage[mathlines]{lineno}

\title{The energy spectrum of ultra-high energy cosmic rays measured at the Pierre Auger Observatory and at the Telescope Array}

\ShortTitle{Auger and TA UHECR energy spectrum}

\author{\speaker{Olivier Deligny}$^a$ for the Pierre Auger${^b}$~and Telescope Array${^c}$ Collaborations\footnote{for collaboration lists see PoS(ICRC2019)1177}\\
\llap{$^a$}CNRS/IN2P3 - IPN Orsay, France\\
\llap{$^b$}Observatorio Pierre Auger, Av.\ San Mart\'in Norte 304, 5613 Malarg\"ue, Argentina\\
E-mail: \href{mailto:auger_spokespersons@fnal.gov}{\rm auger\_spokespersons@fnal.gov}\\
Full author list: \href{http://www.auger.org/archive/authors_icrc_2019.html}{\rm http://www.auger.org/archive/authors\_2019.html}\\
\llap{$^c$}Telescope Array Project, 201 James Fletcher Bldg, 115 S. 1400 East, Salt Lake City, UT 84112-0830, USA\\
E-mail: \href{mailto:ta-icrc@cosmic.utah.edu}{\rm ta-icrc@cosmic.utah.edu}\\
Full author list: \href{http://www.telescopearray.org/index.php/research/publications/conference-proceedings}{\rm http://www.telescopearray.org/index.php/research/publications/conference-proceedings}}

\abstract{The energy spectrum of ultra-high energy cosmic rays (UHECRs) provides essential information on the most energetic phenomena in the Universe. Beyond EeV energies, the Telescope Array and Pierre Auger Observatory have the largest exposures to UHECRs ever accumulated in the Northern and Southern hemispheres, respectively. The results show independently a steepening of the energy spectrum above a few tens of EeV. However, the comparison of the spectra shows differences that are not explainable in terms of an overall uncertainty on the energy scale used to reconstruct the extensive air showers. The differences are also observed in the region of the sky covered by both instruments, where the spectra should be in agreement within uncertainties when directional-exposure effects are accounted for. For this contribution, a working group from both collaborations examined these differences considering the energy-dependent systematic uncertainties. A special focus is given to the characterization of the spectral features, which provide an accurate tool to enhance our understanding of the comparisons.}

\FullConference{36th International Cosmic Ray Conference -ICRC2019-\\
		July 24th - August 1st, 2019\\
		Madison, WI, U.S.A.}

\begin{document}
\setcounter{page}{2}

\section{\label{section:intro} Introduction}

Cosmic rays compose less than one particle out of ten million in the interstellar gas. Still, their average energy density is similar to that of the gas. A small proportion of particles has therefore appropriated a substantial part of the available energy. The study of the energy spectrum of cosmic rays, providing the differential intensity (flux per steradian) of cosmic protons and nuclei as a function of energy, is thus one of the cornerstones of astroparticle physics. 

Because of the very small value of the cosmic-ray intensity at high energies -- less than one particle per km$^2$~yr~sr above 10~EeV -- the construction of giant observatories has been necessary to collect an increased influx of events. The Pierre Auger Observatory, located in the province of Mendoza (Argentina) and covering 3000~km$^2$, has been allowing since 2004 a scrutiny of the UHECR intensity -- except in the northernmost quarter. 
Another scrutiny, mainly of the Northern sky, has been provided by the Telescope Array (TA), located in Utah (USA) and covering 700~km$^2$, operating since 2008. 
These latest-generation experiments have allowed an unprecedented sensitivity in measuring the UHECR energy spectrum. 

\begin{figure}[!t]
  \centering
  \includegraphics[width=0.9\textwidth]{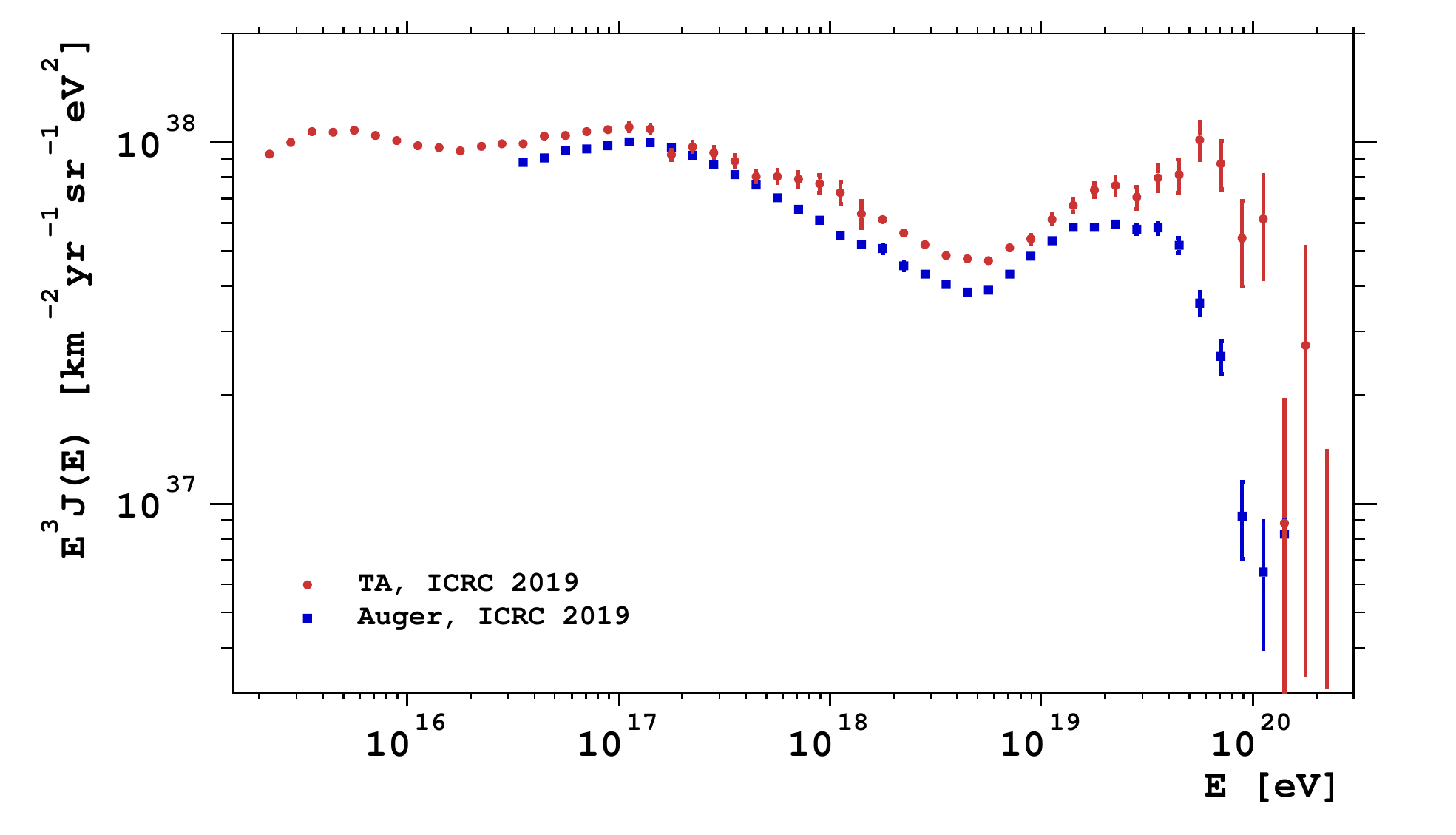}
  \caption{ICRC 2019 energy spectra of the Pierre Auger Observatory and the Telescope Array scaled by $E^3$. In each experiment, data 
  of different detection techniques are combined to obtain the spectrum over a wide energy range.}
  \label{fig:icrc19spectra}
\end{figure}

In this joint contribution, we review the different energy spectrum measurements made at these observatories in the last decade in the quest to decipher the UHECR origin. Both observatories are hybrid cosmic-ray detectors that consist of fluorescence telescopes overviewing an array of surface detectors (SD). The fluorescence detectors (FD) provide an accurate determination of the cosmic-ray energies by measuring the longitudinal developments of the extensive air showers in a nearly calorimetric manner. Their duty cycle is however limited to about 15\%. By contrast, the SD duty cycle is quasi-permanent, allowing for a large and uniform exposure. It is thus advantageous for both Auger and TA to use their SD arrays to measure the energy spectrum at the highest energies, by propagating the FD energy scale to the SD thanks to a subset of events simultaneously detected by both the FD and SD.

In addition to the detectors dedicated to the highest energies, low-energy extensions have allowed for lowering significantly the thresholds. In particular, at lower energies where the fluorescence light produced by extensive air showers becomes too weak to be detected, showers aimed more directly at the FD telescopes can still be detected efficiently through the air-Cherenkov light emitted by the charged particles. Such showers can be reliably reconstructed by making use of the so-called profile-constrained geometry fit, initially designed to reconstruct fluorescence-dominated monocular events at the HiRes experiment~\cite{hires_pcgf}. 

Combining the different detection techniques available, the Auger and TA energy spectra reported at this conference, scaled by $E^3$, are shown in figure~\ref{fig:icrc19spectra}. At first glance, modulo a global offset, consistent spectral features are observed. More detailed comparisons are the object of this contribution, focusing on the energy range below $10~$EeV in section~\ref{section:features} and above $10~$EeV in section~\ref{section:fittedspectra}. In section~\ref{section:declination}, we present further comparisons, through the declination dependence of the spectrum above $10~$EeV. Some concluding remarks are given in section~\ref{section:conclusions}. But prior to these studies, we highlight in section~\ref{section:lessons} the main lessons of the previous comparisons carried out in the last decade.

\section{\label{section:lessons} Previous comparisons: lessons from a ten-year endeavor}

Although the techniques for assigning energies to events are nearly the same, there are differences as to how the primary energies are derived at both observatories as well as and differences in the energy spectrum estimates. Currently, systematic uncertainties in the energy scale of both experiments amount to about 14\% (Auger) and 21\% (TA). Uncovering the sources of systematic uncertainties in the relative energy scale is ultimately necessary to understand the differences in the energy spectra, in particular at the highest energies. To this aim, a joint working group has been formed between both collaborations in December 2011, in view of the UHECR 2012 meeting held at CERN. Throughout the series of UHECR conferences that followed, notable results of this joint effort have been obtained among which the three important ones are summarized below:
\begin{itemize}
\item The difference in normalization of the spectra is attribuable to a mismatch in the relative energy scales of the order of 10\%. Such an amount is naturally within the budget of systematic uncertainties in the energy scale, in particular of the parameters external to the reconstruction such as the invisible energy and the fluorescence yield. For instance, the shift in energies obtained by applying the fluorescence yield adopted by the TA (Auger) Collaboration to the Auger (TA) reconstruction amounts to $\simeq +12\%$ ($\simeq -14\%$). 
\begin{figure}[!t]
  \centering
  \includegraphics[width=0.49\textwidth]{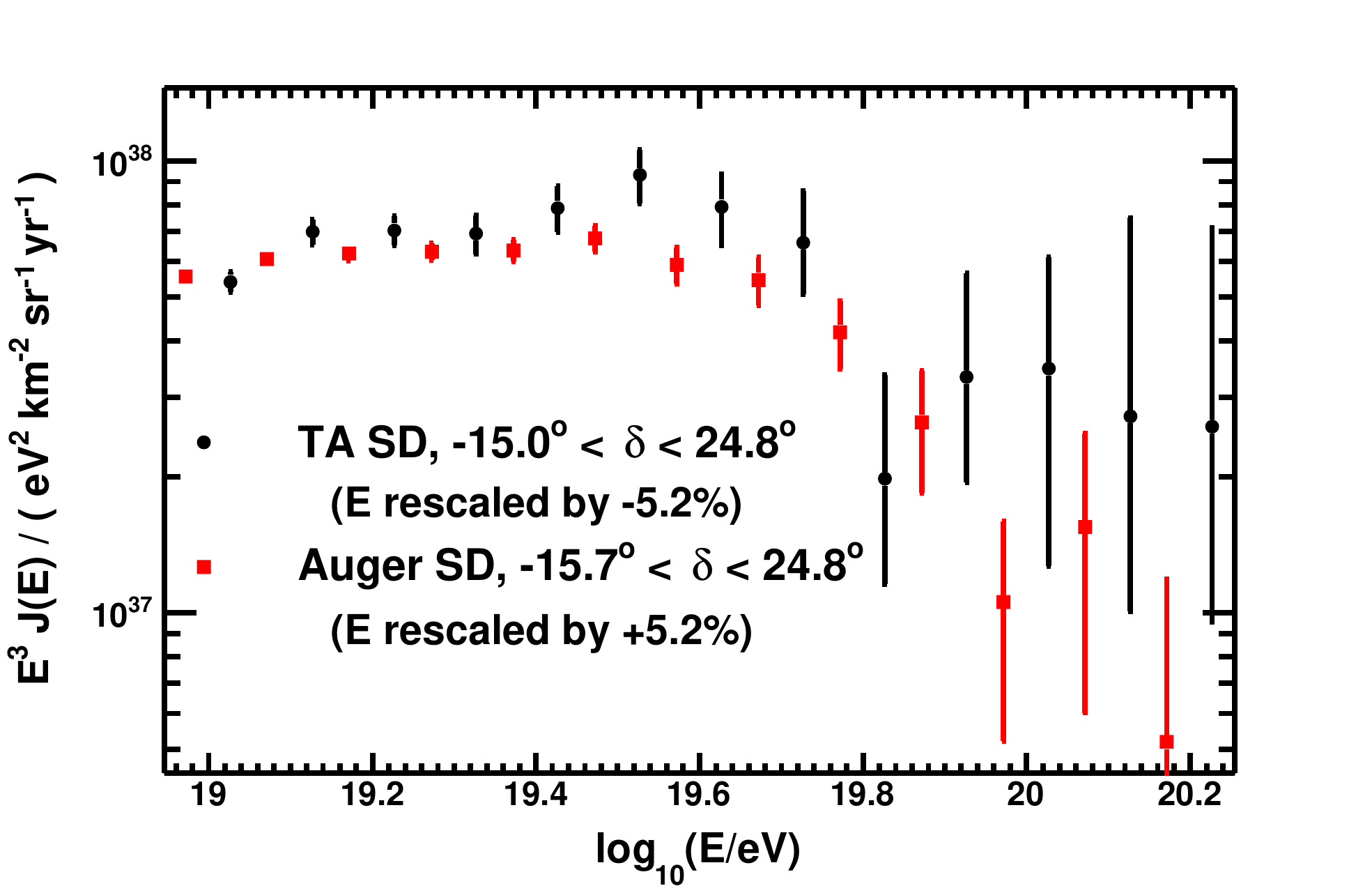}
  \includegraphics[width=0.49\textwidth]{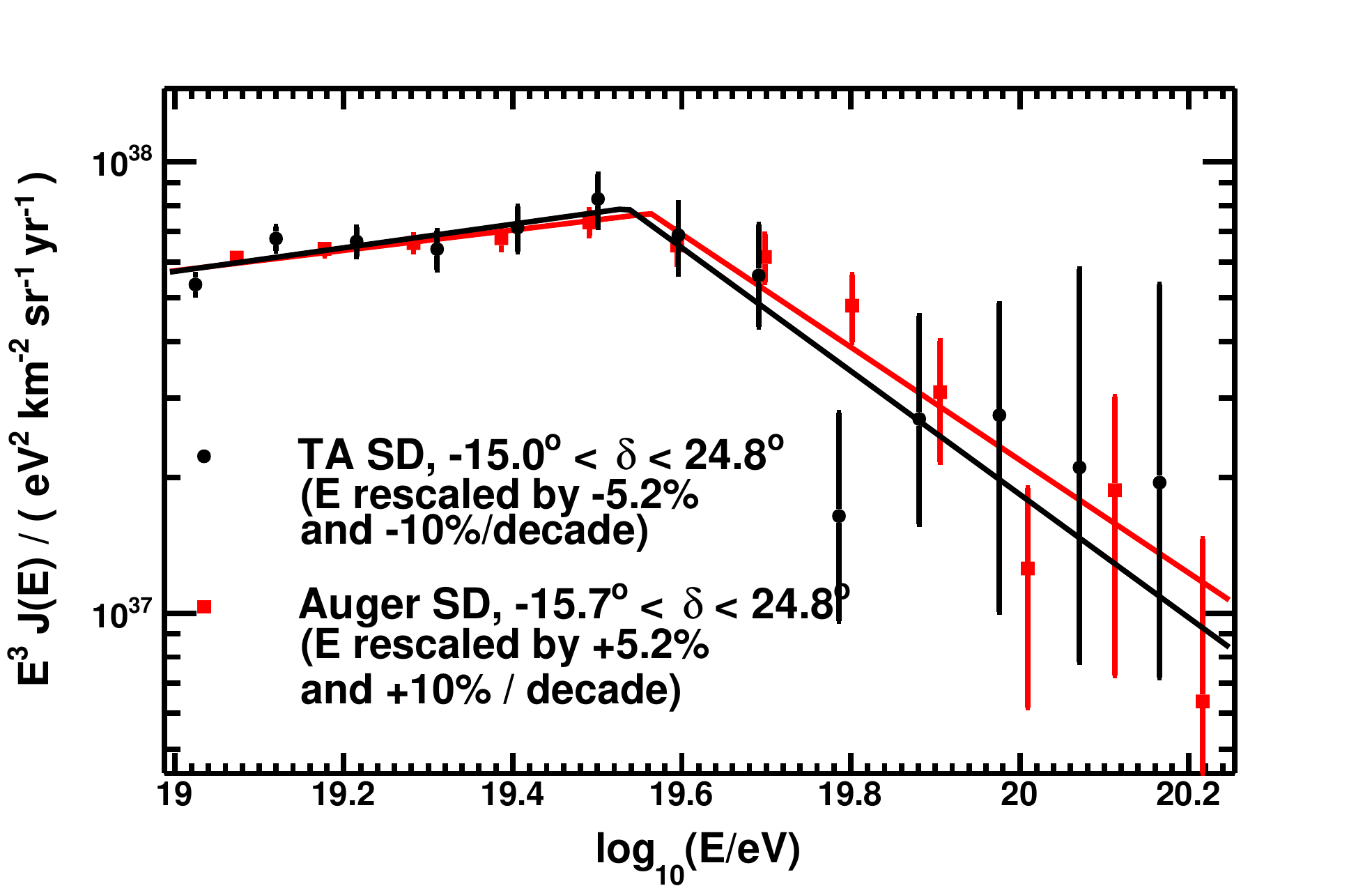}
  \caption{Left: TA and Auger $E^3$-scaled spectra above $10$~EeV in the common declination band after applying overall +5.2\% (Auger) and -5.2\% (TA) energy scale corrections. Right: same after applying additional energy-dependent energy corrections to TA and Auger. From~\cite{ivanov_uhecr2018}.}
  \label{fig:spectrum_commonband}
\end{figure}
\item After applying a global renormalization of the respective energy scales by $+5.2\%$ for Auger and $-5.2\%$ for TA, consistent spectra are obtained in the ankle-energy region, up to $\simeq 10~$EeV. However, persistent differences remain above $\simeq 10~$EeV. To disentangle anisotropy from systematic effects, a detailed scrutiny of the spectra in the declination range commonly observed has been carried out, removing the distortions of the respective directional exposures to a possible anisotropic intensity~\cite{verzi_uhecr2016,ivanov_icrc2017}. The resulting spectra are shown in figure~\ref{fig:spectrum_commonband}-left: although the cutoff energies become statistically consistent, there are differences in the shape of the spectra. A further empirical, energy-dependent systematic shift of +10\% (-10\%) per decade for Auger (TA) is required to bring spectra in agreement, as observed in the right panel of the figure. 
\item A comprehensive search for energy-dependent systematics on the energies has been undertaken~\cite{ivanov_uhecr2018}. This includes the adopted fluorescence yield, the uncertainties in the correction factor for the invisible energy, the influence of the atmospheric transmission used in the reconstruction, the uncertainties in the energy calibration, and the residual bias of the SD energies once calibrated with the FD ones. The net result of this comprehensive review is that non-linearities in the $[10 - 100]~$EeV decade amount to $\pm 3\%$ for Auger and $(-0.3\pm9)\%$ for TA, which are below the required $+10\%$ per decade (Auger) and $-10\%$ per decade (TA). 
\end{itemize}
Note also that the TA energies have been shown to depend only very little (within 3\%) on the method used to convert the SD shower size into energy, be it by making use of the constant intensity cut (CIC) method to infer the attenuation curve and to calibrate the obtained energy estimator with FD energies, or else making use of a lookup table derived from the CORSIKA QGSJET-II.3 proton Monte-Carlo simulations, with final energies globally rescaled to the FD ones~\cite{ivanov_uhecr2018}. Besides, neglecting the energy dependence of the attenuation curve as inferred from the CIC method was shown to lead to a $\pm 2$\% effect on the energies at the Auger Observatory~\cite{ivanov_uhecr2018}.

\section{\label{section:features} Comparisons of the low-energy spectral features}

A customary scheme to describe the energy spectra shown in figure~\ref{fig:icrc19spectra} is through the fit of a series of broken power laws. The power-law exponents and break energies are the spectral features capturing the rate of fall of the intensity with energy. The overlap in the energy range from $3\times 10^{16}~$eV to the highest energies between the Auger and TA measurements allows for comparing the gradual fall-off of the intensity in the ``second knee'' (or ``iron knee'')-to-suppression region, spanning four orders of magnitude in energy. 

A first energy break is obtained at $(0.11\pm0.01)$~EeV for TA (the uncertainty is the statistical one), and at $(0.15\pm0.02)$~EeV for Auger (the uncertainty combines the statistical and systematic errors). This energy break marks a steepening of the spectral index going from $2.920\pm0.008$ to $3.15\pm0.03$ for TA, and from $2.92\pm0.05$ to $3.27\pm0.05$ for Auger. There is a good statistical agreement between these reported values, the largest difference being at the $2.1\sigma$ level for the second spectral index. A widespread view to interpret this steepening is that the intensity of each individual nuclear component of the bulk of Galactic cosmic rays falls off steeply at a magnetic rigidity near $3~$PV, as a consequence of the difficulty for diffusive shock acceleration in supernova remnants to produce particles beyond this limit. The second knee would thus mark the rapid fall-off of the Galactic intensity for iron nuclei. 

Both TA and Auger spectra below $\simeq0.1~$EeV are built by making use of the detection of showers observed with the air-Cherenkov light emitted by the charged particles. The external input to the reconstruction is thus exclusively the invisible-energy correction, which is mass-dependent. The comparison effort just started at the time of the conference will thus benefit from a detailed scrutiny of the different inputs used to model this correction. 

The rate of fall becomes less steep again at the ankle energy, measured at $(4.9\pm0.1)$~EeV for TA and at $(6.2\pm0.9)$~EeV for Auger. More detailed comparisons of the spectra in the ankle-to-suppression region are given in the next section.

\section{\label{section:fittedspectra} Fit models of the differential energy spectra}

As described in section~\ref{section:lessons}, on top of a global rescaling of the energies, an additional non-linearity in the relative Auger/TA energy scale has been previously invoked to bring the spectra in agreement in the ankle-to-suppression region. In this section, we aim at making use of the fit functions of the differential spectra to infer the energy-dependent translation of energies needed to get consistent spectra. Two models of fit functions are used: a single broken power law, and a double broken power law. 

Let $J_1(E)$ be the fit function of the differential spectrum 1, and $J_2(E)$ that of the differential spectrum 2. The local energy shift, denoted as $b(E)$, that allows for switching from $J_2(E)$ to $J_1(E)$ is inferred from the condition $J_1(E+b(E))=J_2(E)$, which can be rewritten in terms of a Taylor series as
$J_1(E)-J_2(E)+\sum_{k\geq1}\frac{[b(E)]^k}{k!}\frac{d^kJ_1(E)}{dE^k}=0.$
This latter equation can be solved numerically to determine the function $b(E)$. Note that the $b(E)$ function obtained from this procedure is not to be directly compared to the correction factor of the energy scale discussed in section~\ref{section:lessons}. This is because a change of spectrum function under a change of energy scale is subject to the jacobian transformation of the energy scale, $J_1(E_1)dE_1=J_2(E_2)dE_2$. This latter approach leads however to more complex equations to solve, left for future studies. The approach adopted here is useful to probe the energy ranges in which a single linear energy shift is required to bring spectra in agreement, and to probe the energy ranges in which such shifts require terms beyond the linear one. 

The uncertainties on $b(E)$ are obtained by propagating the variance/covariance matrix of the fitted parameters defining $J_1(E)$ and $J_2(E)$. The analysis is performed twice, by considering alternatively TA or Auger for the spectrum function $J_1(E)$. Note that there are unavoidable bin-to-bin correlations in the output of this analysis for the obtained $b(E)$. 

\begin{figure}[!t]
  \centering
  \includegraphics[width=0.49\textwidth]{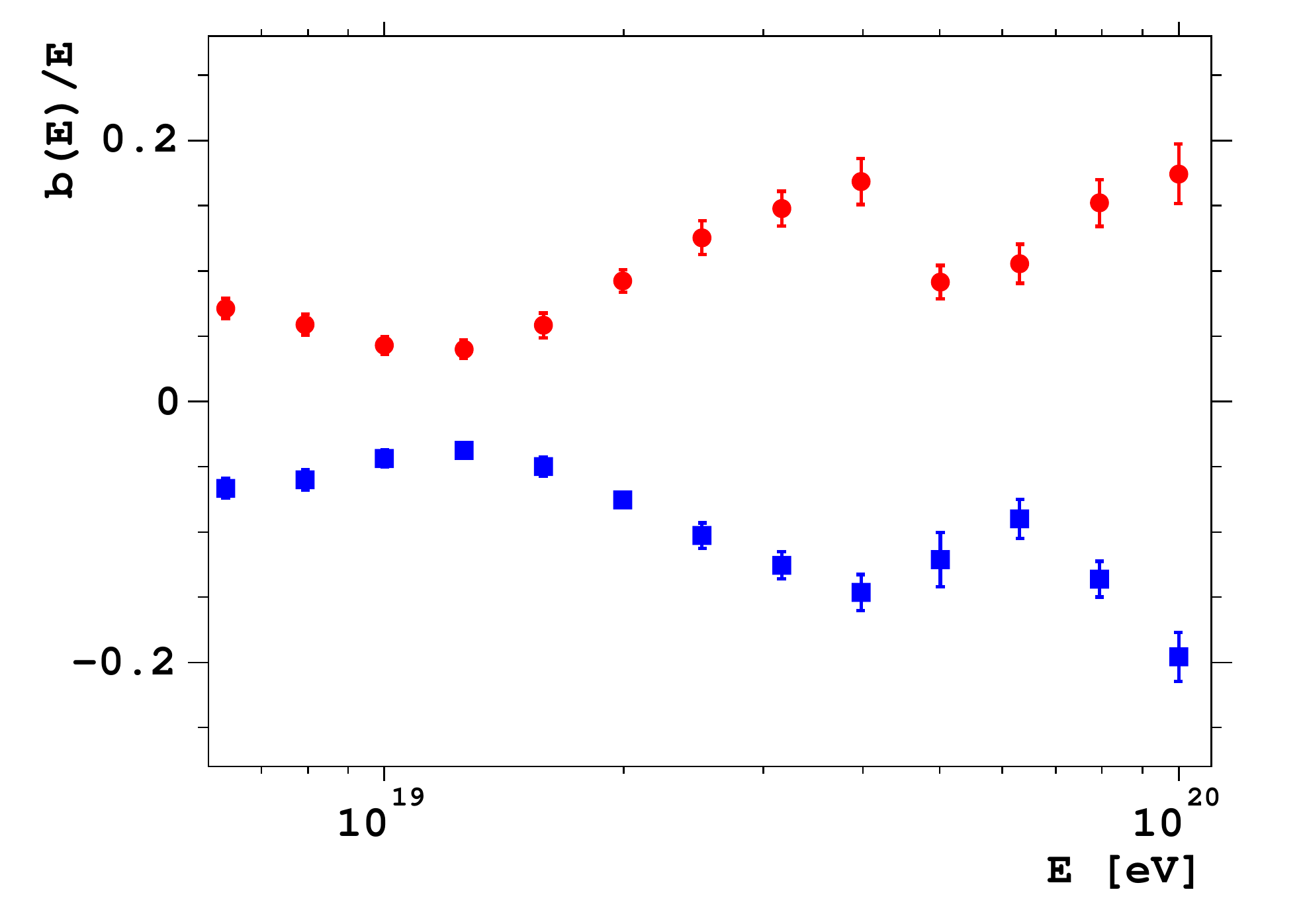}
  \includegraphics[width=0.49\textwidth]{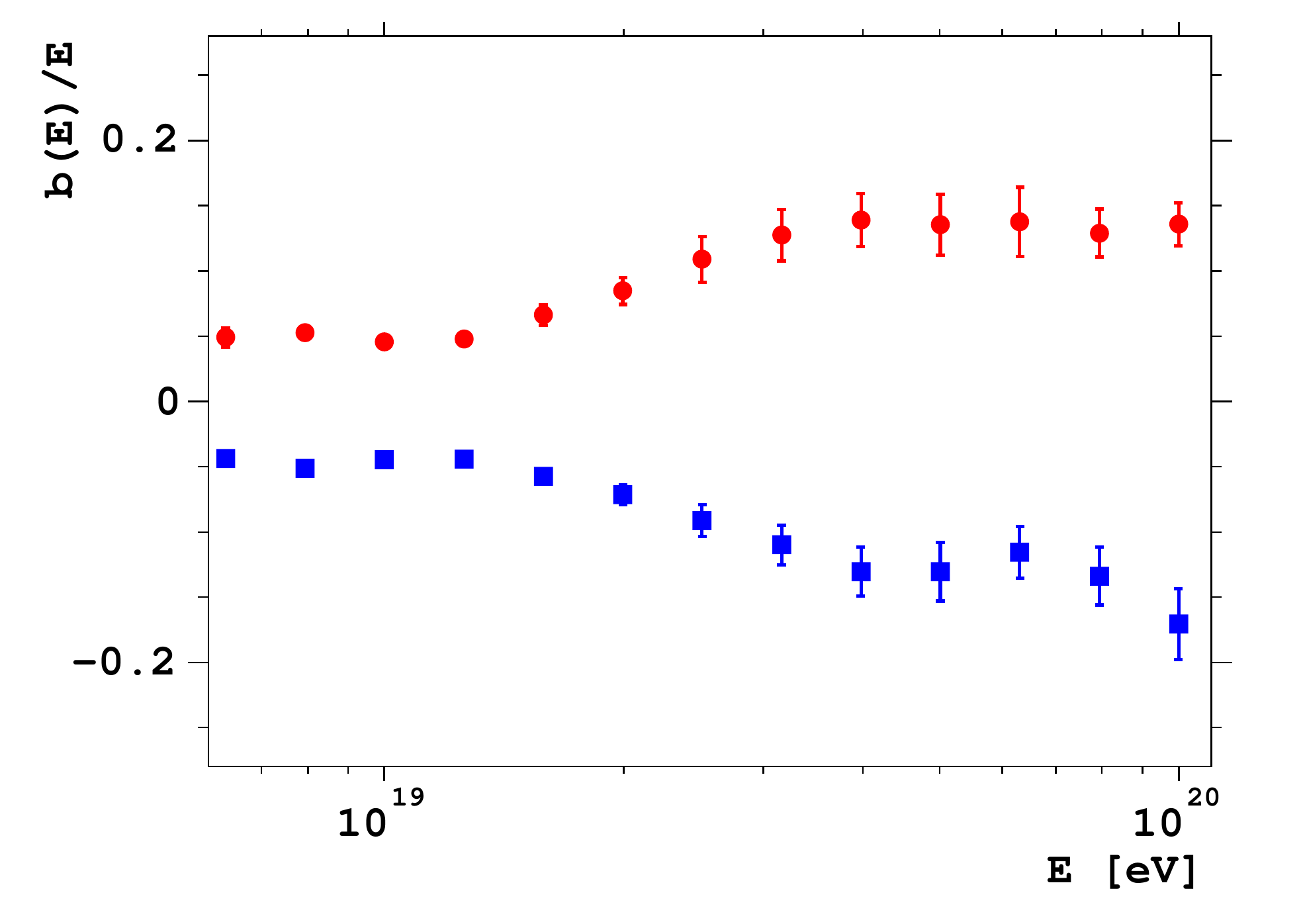}
  \includegraphics[width=0.49\textwidth]{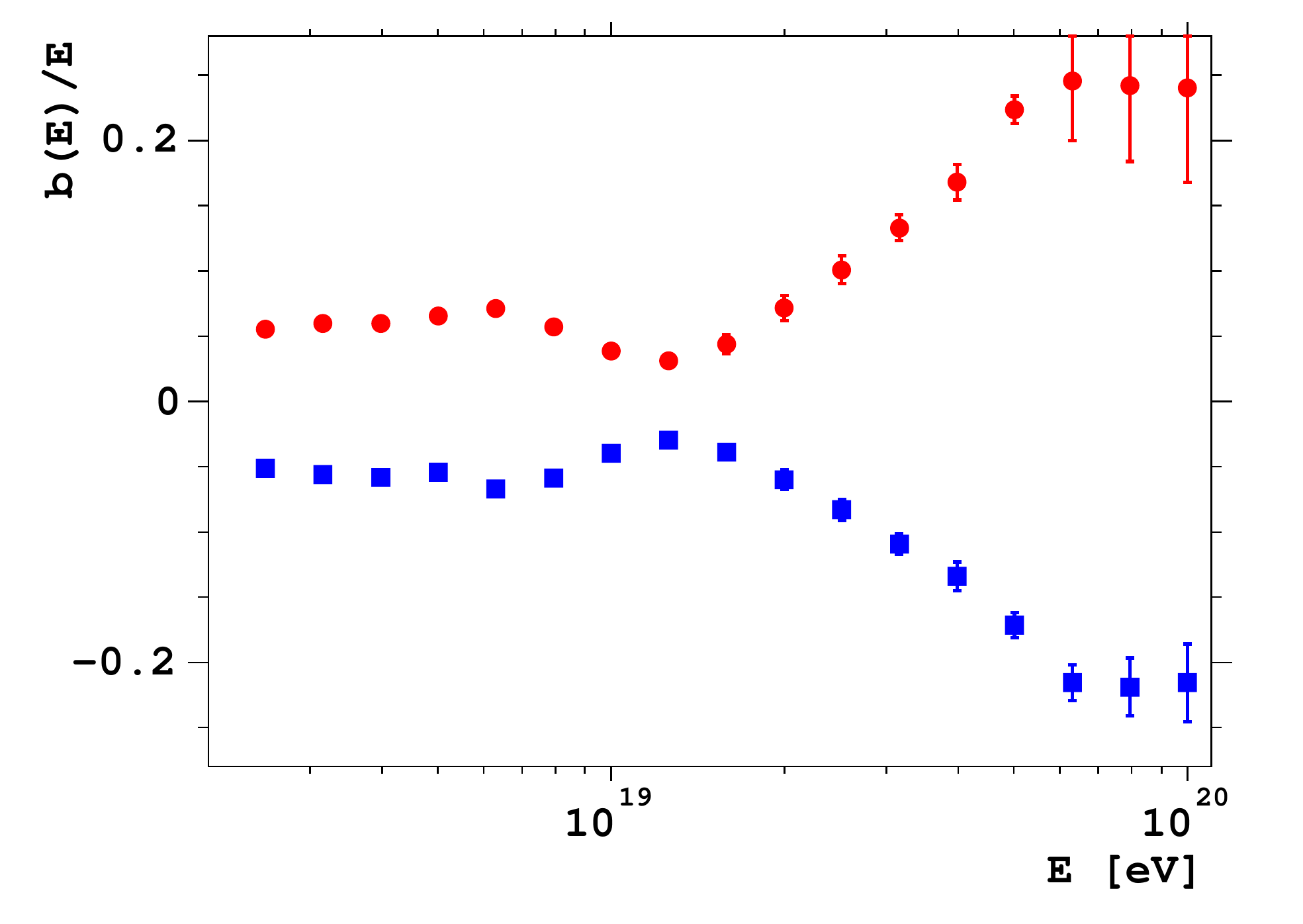}
  \includegraphics[width=0.49\textwidth]{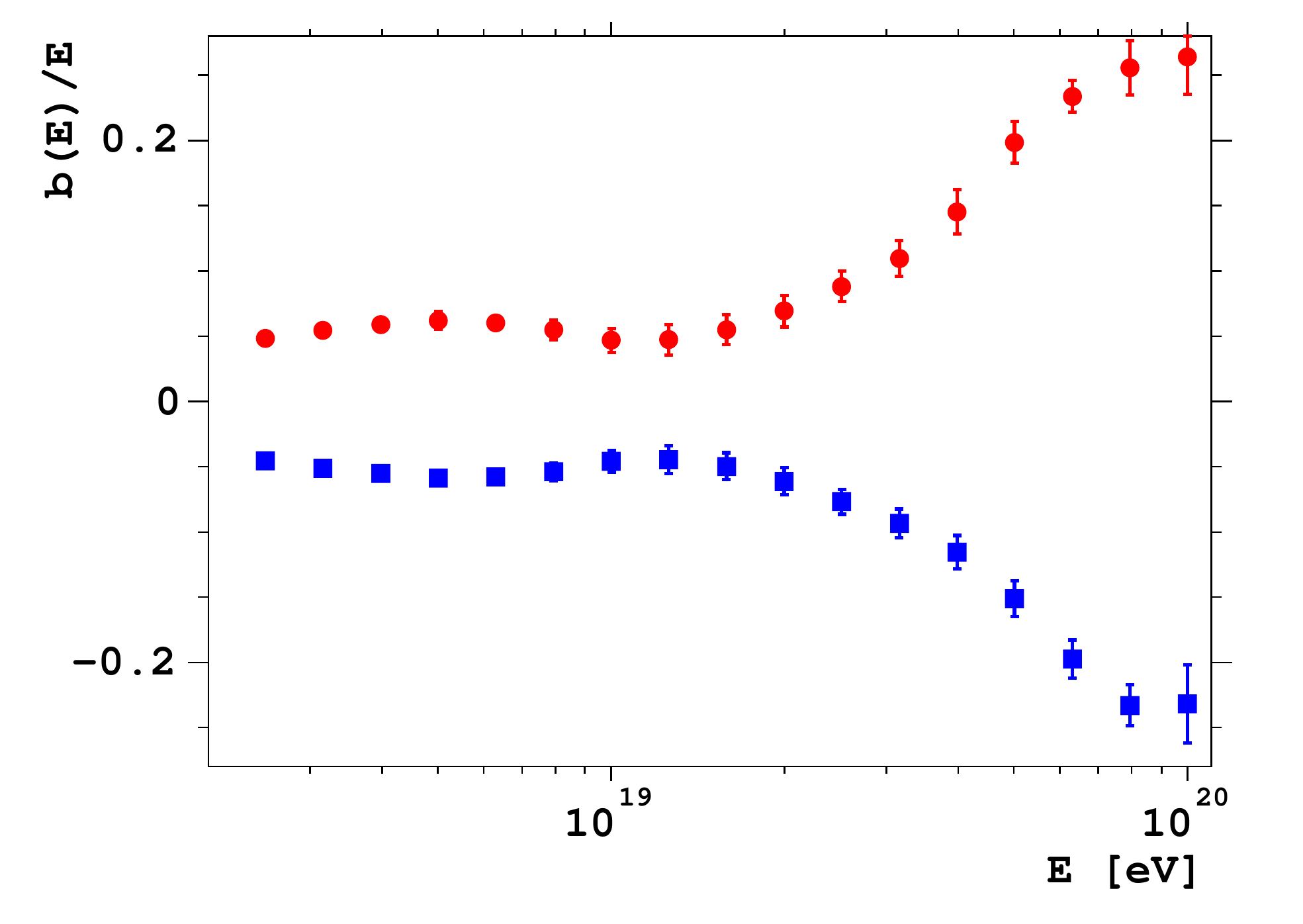}
  \caption{Energy shift term needed to bring the fitted differential spectra in agreement. Left: single broken power law for the underlying spectrum. Right: double broken power law for the underlying spectrum. Top: analysis in the common declination band, using an energy spectrum estimator free of directional-exposure distortions of anisotropies. Bottom: analysis in the whole fields of view of each observatory.}
  \label{fig:b}
\end{figure}

Results are displayed in figure~\ref{fig:b}, considering a single broken power law for the fit function in the left panels, and a double broken power law in the right ones. The top panels are obtained when restricting the analysis to the declination band commonly observed, with blue points referring to $J_1(E)$ for Auger and red ones to $J_1(E)$ for TA. In the case of the double broken power-law fit function of the spectra, it is observed that a single linear energy shift (that is, a constant $b(E)/E$ factor) is enough to bring spectra in agreement from 6 to $\simeq 15~$EeV, and that a single non-linear term is required from $\simeq 15~$EeV to the suppression region. By contrast, the situation gets more complex in the case of the single broken power-law fit function. In particular, a linear energy shift never allows for bringing spectra in agreement in that case. The same conclusion holds when applying the analysis to the spectra obtained in the whole fields of view of each observatory, as shown in the bottom panels of figure~\ref{fig:b}. 

From these comparisons, a single non-linear term in the relative energies, as inferred from previous work, prefers a double broken power-law fit function for the underlying differential spectrum.

\section{\label{section:declination} Declination dependences}

\begin{figure}[!t]
  \centering
  \includegraphics[width=0.49\textwidth]{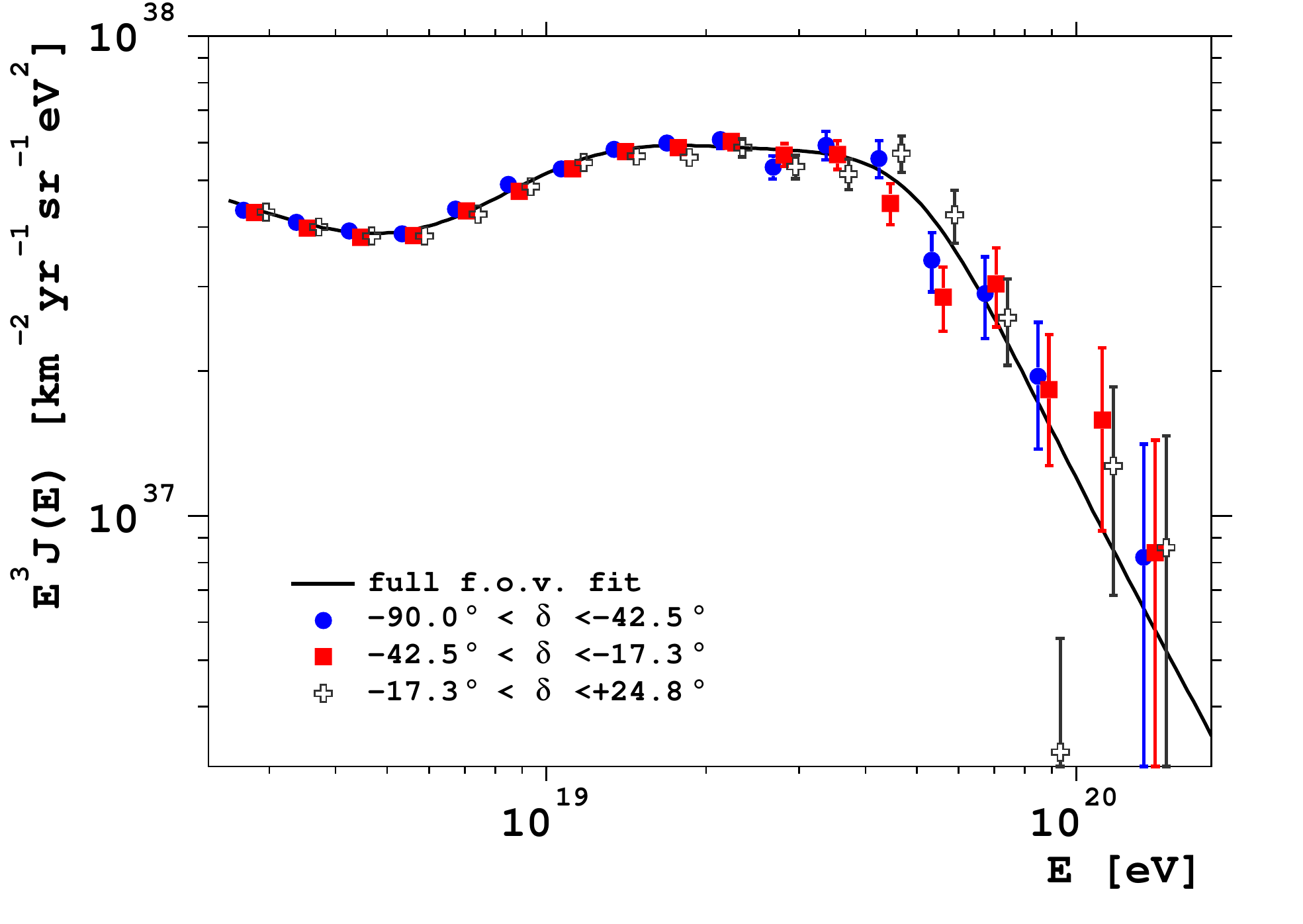}
  \includegraphics[width=0.49\textwidth]{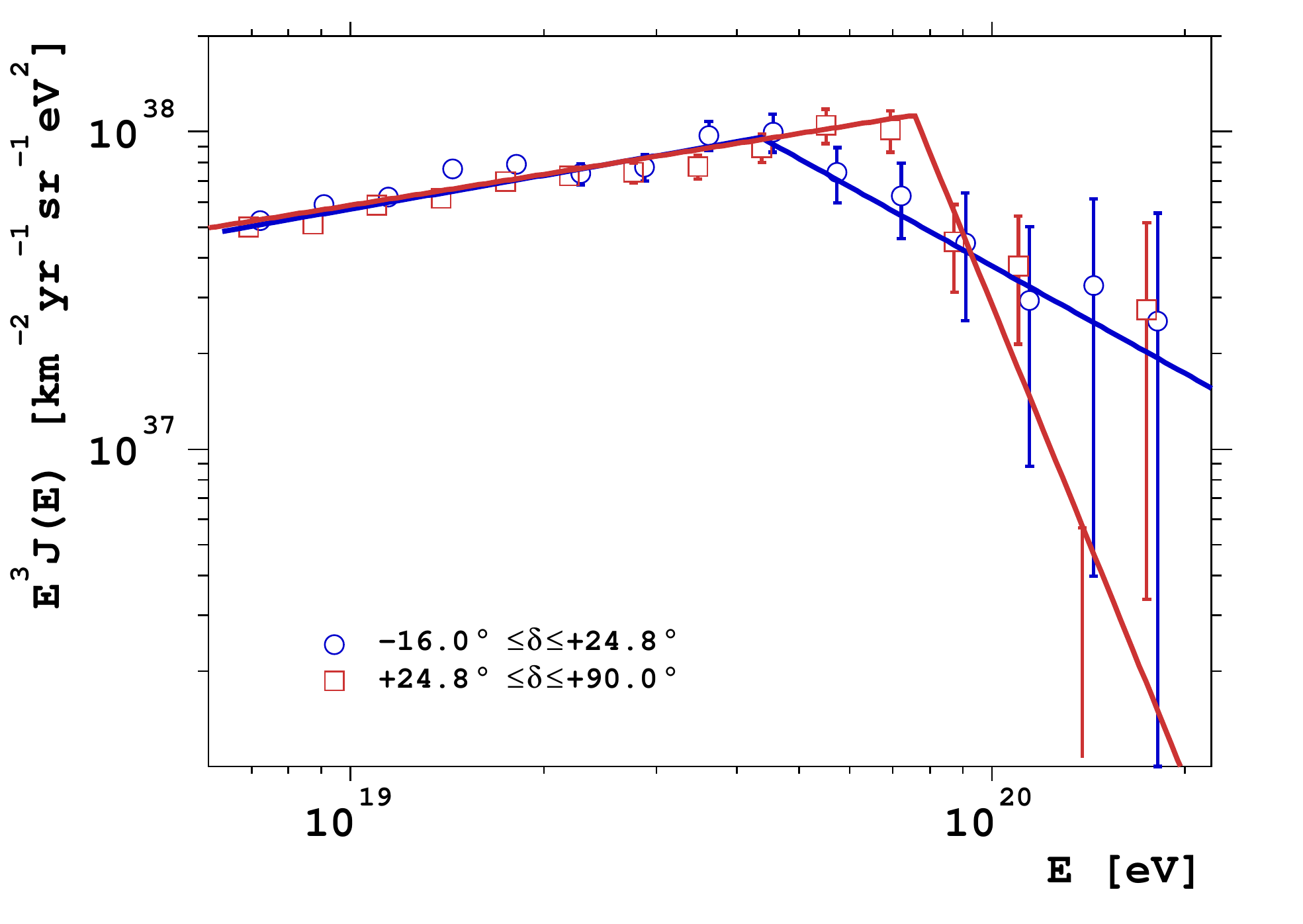}
  \caption{$E^3$-scaled energy spectra in different declination ranges. Left: Auger. Right: TA.}
  \label{fig:declination}
\end{figure}

The wide range of declinations covered by each observatory allows a search for dependences of the energy spectrum on declination. In the case of Auger, the resulting spectra (scaled by $E^3$) are shown in the left panel of figure~\ref{fig:declination} (an artificial shift of $\pm 5\%$ is applied to the energies of the northernmost/southernmost declination spectra to allow for distinguishing the various data points). For reference, the best fit of the overall spectrum is shown as the black line. No obvious dependence with declination is observed over the whole energy range covered. In the case of TA, the $E^3$-scaled spectra are shown in the right panel. It can be seen that the break point (using the broken power-law fit)  for lower declinations $-16^\circ < \delta < 24.8^\circ$ (shown as the red points) occurs at $\log_{10}(E/\mathrm{eV}) = 19.59 \pm 0.06$, while for $24.8^\circ < \delta < 90^\circ$ (shown as the blue points), the break point occurs at $\log_{10}(E/\mathrm{eV}) = 19.85 \pm 0.03$. Stringent tests of the detector systematic uncertainties show the absence of declination-dependent bias of the energies so that the effect must be astrophysical~\cite{ta_icrc2019}. The global significance for this effect has been estimated to be 4.3~$\sigma$~\cite{ta_icrc2019}.

\begin{figure}[!t]
  \centering
  \includegraphics[width=0.7\textwidth]{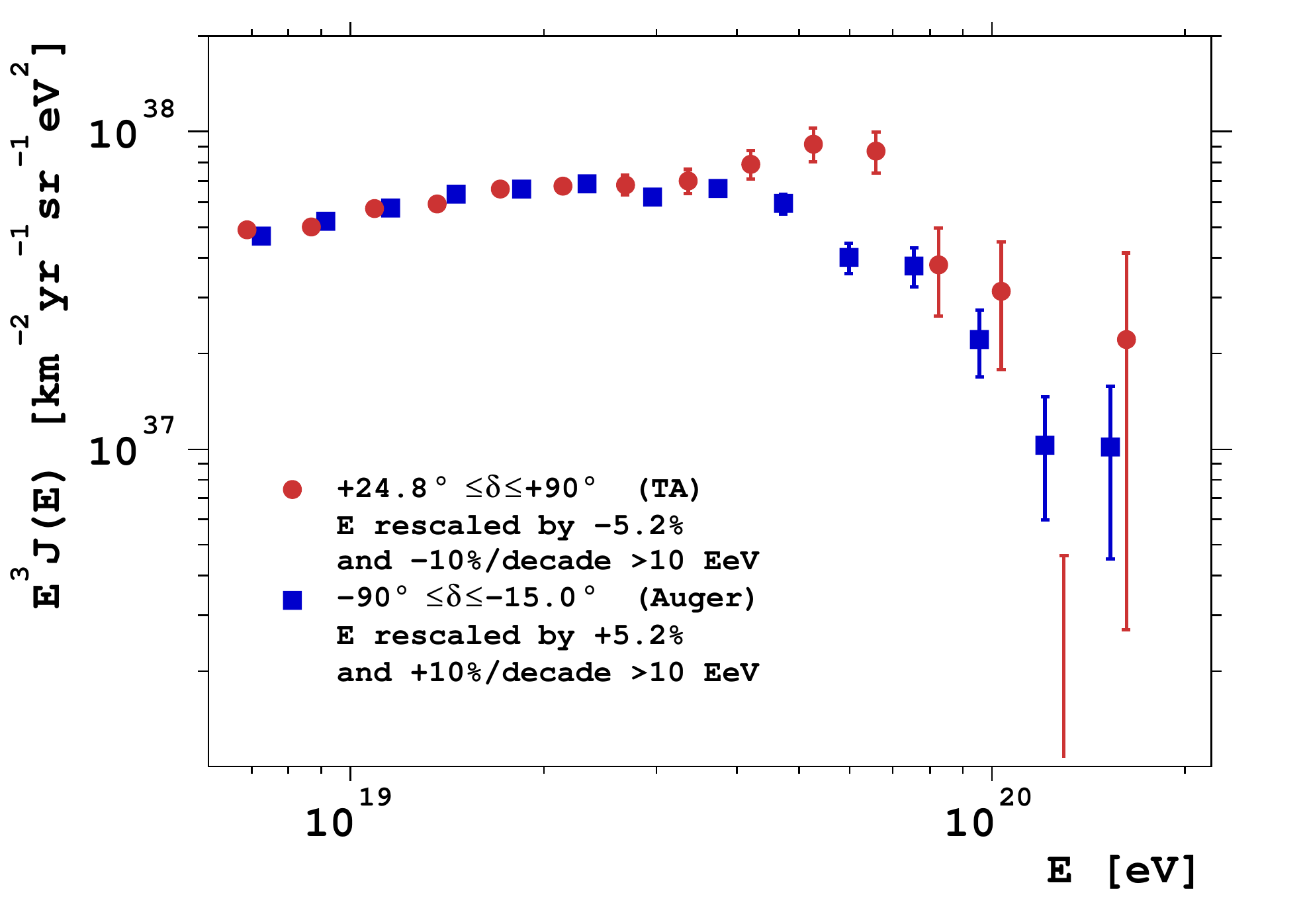}
  \caption{$E^3$-scaled energy spectrum in two declination ranges, namely $\delta\geq+24.8^\circ$ (red points, TA) and $\delta\leq-15.0^\circ$ (blue points, Auger). The energies are shifted to get spectra in agreement in the common declination band.}
  \label{fig:declination2}
\end{figure}

In this joint work, we provide further comparisons related to declination dependences of the spectrum. We use the energy rescaling that allows for bringing spectra in agreement in the common declination band to compare the northernmost declination band ($\delta\geq+24.8^\circ$ in the TA field of view, shown as the red points in figure~\ref{fig:declination2}) to the southernmost one ($\delta\leq-15.0^\circ$ in the Auger field of view, shown as the blue points). An excess of intensity is observed in the northernmost declination range (TA) around the break energy.

\section{\label{section:conclusions} Conclusions}

We have reviewed and compared the results of the energy spectra measured at the Pierre Auger Observatory and at the Telescope Array. From these comparisons, a few concluding remarks are in order:
\begin{itemize}
\item There is good agreement between the spectral features in the second knee-to-ankle energy range, modulo a rescaling factor of the energies;
\item The global rescaling of energies required to bring the SD spectra in agreement from the ankle energy to $\simeq 10~$EeV is attribuable to the different fluorescence yields used at each observatory;
\item On top of a global rescaling of energies, a non-linearity is needed to bring spectra in agreement in the range of common declinations;
\item A single non-linear term in the relative energies prefers a double broken power-law fit function for the underlying differential spectrum.
\end{itemize}
The sources of the non-linearity have not been identified, yet. Further studies of the relative systematic uncertainties between TA and Auger will benefit from:
\begin{itemize}
\item The second phase of the deployment of Auger detectors at the TA site for cross-checking the response of the two different detection techniques to the same extensive air showers;
\item The deployment of scintillators at the Auger site in the context of its upgrade, that will allow for mimicking the TA scheme to some extent;  
\item The reduction of the statistical uncertainties with the future operation of TA$\times4$ and the continuous Auger data taking.
\end{itemize}

\end{document}